\documentclass[preprint,12pt,longnamesfirst]{aastex}

\usepackage{latexsym}

\newcommand {\ha}{H$\alpha$}
\newcommand {\kms}{km~s$^{-1}$}
\shorttitle{Diffuse X-Ray Emission from M\,17}
\shortauthors{Dunne et al.}

\begin{document}

\title{Diffuse X-Ray Emission from the Quiescent Superbubble M\,17,
the Omega Nebula}

\author{Bryan C. Dunne\altaffilmark{1}, You-Hua Chu\altaffilmark{1}, 
C.-H. Rosie Chen\altaffilmark{1}, Justin D. Lowry\altaffilmark{1}, 
Leisa Townsley\altaffilmark{2}, Robert A. Gruendl\altaffilmark{1},
Mart\'{\i}n A. Guerrero\altaffilmark{1}, and Margarita
Rosado\altaffilmark{3}}

\email{carolan@astro.uiuc.edu}

\altaffiltext{1}{Department of Astronomy, University of Illinois, 1002
West Green Street, Urbana, IL 61801}

\altaffiltext{2}{Department of Astronomy \& Astrophysics, Pennsylvania
State University, 525 Davey Laboratory, University Park, PA 16802}

\altaffiltext{3}{Instituto de Astronom\'{\i}a, Universidad Nacional
Aut\'onoma de M\'exico, Apdo. Postal 70-264, 04510 M\'exico, D.F.,
M\'exico}

\begin{abstract}

The emission nebula M\,17 contains a young $\sim$1~Myr-old open
cluster; the winds from the OB stars of this cluster have blown a
superbubble around the cluster.  {\it ROSAT} observations of M\,17
detected diffuse X-ray emission peaking at the cluster and filling the
superbubble interior.  The young age of the cluster suggests that no
supernovae have yet occurred in M\,17; therefore, it provides a rare
opportunity to study hot gas energized solely by shocked stellar winds
in a quiescent superbubble.  We have analyzed the diffuse X-ray
emission from M\,17, and compared the observed X-ray luminosity of
$\sim$2.5$\times$10$^{33}$~ergs~s$^{-1}$ and the hot gas temperature
of $\sim$8.5$\times$10$^6$~K and mass of $\sim$1~M$_\odot$ to model
predictions.  We find that bubble models with heat conduction
overpredict the X-ray luminosity by two orders of magnitude; the
strong magnetic fields in M\,17, as measured from \ion{H}{1} Zeeman
observations, have most likely inhibited heat conduction and
associated mass evaporation.  Bubble models without heat conduction
can explain the X-ray properties of M\,17, but only if cold nebular
gas can be dynamically mixed into the hot bubble interior and the
stellar winds are clumpy with mass-loss rates reduced by a factor of
$\ge$3.  Future models of the M\,17 superbubble must take into account
the large-scale density gradient, small-scale clumpiness, and strong
magnetic field in the ambient interstellar medium.

\end{abstract}

%\keywords{ISM: bubbles --- ISM: HII regions --- ISM: individual (M17) ---
%stars: early-type --- stars: winds, outflows}

\clearpage
\section{Introduction}
\label{sec:Intro}

Massive stars dynamically interact with the ambient interstellar
medium (ISM) via their fast stellar winds and supernova ejecta.  OB
associations, with their large concentrations of massive stars,
provide an excellent laboratory to study these interactions.  The
combined actions of the stellar winds and the supernovae from the
massive stars in OB associations sweep up the ambient ISM to form
expanding shells called superbubbles \citep{Bruh80}.  The physical
structure of a superbubble is very similar to that of a bubble blown
by the stellar wind of an isolated massive star, as modeled by
\citet{CMW75} and \citet{Weaver77}.

Theoretically, an interstellar bubble consists of a shell of swept-up
ISM with its interior filled by shocked fast wind at temperatures of
10$^{6}$--10$^{8}$~K.  There are two basic types of models for
wind-blown bubbles: energy-conserving and momentum-conserving.  In the
former, the shocked stellar wind is separated from the swept-up
interstellar shell by a contact discontinuity, where heat conduction
and mass evaporation may take place.  The expansion of the shell is
driven by the pressure of the hot interior gas \citep{DdV72,CMW75}.
In the momentum conserving bubbles, the fast stellar winds impinge on
the swept-up shell directly, and the expansion is driven by the
momentum of the fast stellar wind \citep{Aved72,SSW75}.

One significant difference between a superbubble and a single star
bubble is the possibility that supernovae may occur inside a
superbubble and introduce significant perturbations in the surface
brightness and characteristic temperature of the X-ray emission,
especially if a supernova explodes near the dense shell \citep{MMM88}.
This intermittent X-ray brightening has been observed in superbubbles
in the Large Magellanic Cloud (LMC).  Using {\it Einstein} and {\it
ROSAT} observations, \citet{ChuMacLow90} and \citet{Dunne01} have
reported diffuse X-ray emission from a large number of superbubbles in
the LMC, and their X-ray luminosities all exceed the luminosities
expected by Weaver et al.'s bubble model, indicating recent heating by
supernovae.  No LMC superbubbles in a quiescent state, i.e., without
recent supernova heating, have been detected in X-rays by {\it ROSAT}
\citep{Chu95}.  It would be of great interest to detect diffuse X-ray
emission from a quiescent superbubble and compare it to model
expectations, as this could provide a valuable diagnostic of bubble
models.

The emission nebula M\,17 \citep[from the catalog of][$\alpha =
18^{\rm h}21^{\rm m}$, $\delta = -16^{\circ}10'$ (J2000.0), also known
as the Omega Nebula, the Horseshoe Nebula, and NGC\,6618]{Messier1850}
is located on the eastern edge of a massive molecular cloud, M17SW
\citep{LDP74}.  M\,17 exhibits a ``blister-like'' structure with an
overall diameter of $\sim$20$'$--25$'$, or $\sim$10--12~pc at an
adopted distance of 1.6~kpc \citep{Nielbock01}.  The nebula
encompasses an open cluster with a stellar age of $\sim$1~Myr
\citep{Hanson97}.  The open cluster is located on the western side of
the nebula, which borders the molecular cloud.  Arcuate filaments
extend eastward, creating a shell morphology, and suggesting that it
is a young superbubble blown by the OB stars within.  The young age of
the cluster in M\,17 implies that no supernova explosions have
occurred, thus M\,17 provides an ideal setting to study the generation
of hot gas solely by fast stellar winds inside a superbubble at a
quiescent state.

Recent {\it Chandra} observations of M\,17 have revealed diffuse X-ray
emission in the vicinity of the embedded open cluster
\citep{Townsley03}.  As a matter of fact, diffuse X-ray emission from
M\,17 over a more extended area was previously detected in a {\it
ROSAT} observation but was never reported.  We have analyzed this
diffuse X-ray emission from the interior of M\,17 detected by {\it
ROSAT} to determine the physical properties of the hot interior gas,
and considered bubble models with and without heat conduction.  We
find that models with heat conduction produce results with the largest
discrepancy from the observed X-ray luminosity and hot gas temperature
and mass of the superbubble in M\,17.  This paper reports our analysis
of the {\it ROSAT} observations of M\,17 and comparisons to a range of
bubble models.

\section{Observations and Data Reduction}
\label{sec:Data}

\subsection{{\it ROSAT} Archival Data}

We have used an archival {\it ROSAT} Position Sensitive Proportional
Counter (PSPC) observation to study the diffuse X-ray emission from
M\,17 and investigate the physical properties of the hot, shocked gas
interior to the superbubble.  The PSPC is sensitive to X-rays in the
energy range 0.1--2.4 keV and has an energy resolution of $\sim$40\%
at 1 keV, with a field of view of $\sim$2$^\circ$.  Further
information on the PSPC can be found in the \citet{ROSAT91}.

The PSPC observation of M\,17 (sequence number RP500311, PI:
Aschenbach) was obtained on 1993 September 12--13.  It is centered on
the nebula at $\alpha$(J2000) = 18$^{\rm h}$ 21$^{\rm m}$
04\rlap{.}$^{\rm s}$78 and $\delta$(J2000) = $-$16$^{\circ}$ 10$'$
12\farcs0, and has an exposure time of 6.7~ks.  As M\,17 has an
angular size of $\sim$20$'$--25$'$, the diffuse X-ray emission from
the superbubble interior is well contained within the inner window
support ring of the PSPC.  The PSPC data were reduced using standard
routines in the PROS\footnote{PROS/XRAY Data Analysis System --
http://hea-www.harvard.edu/PROS/pros.html} package under the
IRAF\footnote{Image Reduction and Analysis Facility -- IRAF is
distributed by the National Optical Astronomy Observatories, which are
operated by the Association of Universities for Research in Astronomy,
Inc., under cooperative agreement with the National Science
Foundation.} environment.

\subsection{Optical Imaging}

To compare the spatial distribution of the X-ray-emitting gas with
that of the cooler ionized gas in M\,17, we have obtained narrow-band
\ha\ images of this emission nebula.  M\,17 was observed with the
Mount Laguna 1-m telescope on 2002 October 28--31 using a Tektronik 2K
CCD.  A filter with peak transmission at 6563 \AA\ and a FWHM of 20
\AA\ was used to isolate the \ha\ line. Because the 2K CCD has
$\sim$0\farcs 4 pixels and a field-of-view of 13\farcm 6, the entire
nebula could not be observed in a single exposure. Instead a total of
twenty-four 300~s exposures were acquired at 8 positions to span the
nebula.  These exposures were combined to form a mosaic image of the
nebula and to reject cosmic-ray events in the individual images using
the methods outlined in \citet{rg95}.

\section{Analysis of the X-Ray Emission}
\label{sec:Anal}

Significant X-ray emission is detected from M\,17 in the PSPC
observation, as can be seen in Figure~\ref{fig:XrayOpt}.  To
determine the nature and origin of this emission, we have analyzed its
spatial distribution and spectral properties. We have examined the
distribution of the X-ray emission and compared it to the optical
morphology of the emission nebula.  We have also extracted spectra
from the PSPC data and modeled them to determine the physical
conditions of the hot gas.

\subsection{Spatial Distribution of the X-ray Emission}
\label{sec:XrayDist}

\begin{figure}
\begin{center}
\epsscale{1.0}
\plotone{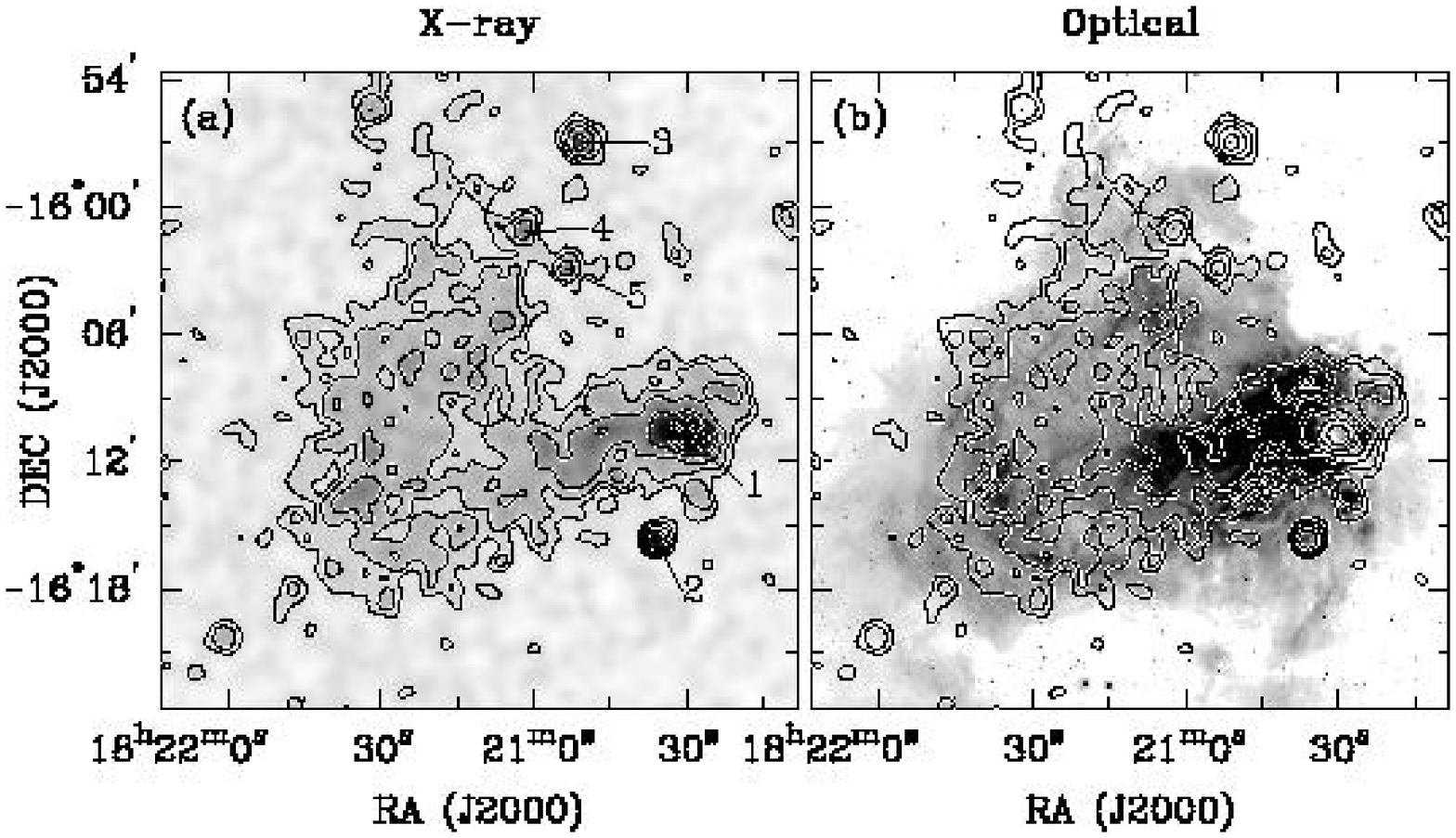}
\caption{M\,17 in X-rays and optical.  (a) shows a smoothed {\it
ROSAT} PSPC X-ray image of M\,17; (b) shows M\,17 in \ha\ emission
overlaid with X-ray contours at 4.5\%, 6.1\%, 10\%, 17\%, 32\%, and
61\% of the peak value.  These contours are also present on the X-ray
image to ensure the clarity of the contours.  Potential point sources
are labeled on the X-ray image: (1) The OB association of M\,17, (2)
1WGA~J1820.6$-$1615, (3) 1WGA~J1820.8$-$1556, (4) 1WGA~J1821.0$-$1600,
and (5) 1WGA~J1820.8$-$1602.
\label{fig:XrayOpt}}
\end{center}
\end{figure}

To study the spatial distribution of the X-ray emission from M\,17,
the data were binned to 5$''$ pixels and then smoothed with a Gaussian
function of $\sigma$ = 4 pixels (see Figure~\ref{fig:XrayOpt}a).  We
have also taken our \ha\ mosaic and overlaid it with X-ray emission
contours to study the extent of X-ray emission within the \ion{H}{2}
region (see Figure~\ref{fig:XrayOpt}b).  Diffuse X-ray emission is
observed to be well confined by the optical nebula.  This diffuse
emission shows no evidence of limb brightening, suggesting that the
interior of M\,17 is centrally filled with hot gas.  The peak of the
X-ray emission is coincident with the center of the open cluster in
M\,17, where {\it Chandra} observations show a large number of point
sources superposed on the diffuse emission \citep{Townsley03}.  This
spatial distribution suggests that the X-ray-emitting hot gas
originates in the open cluster, as would be expected in a bubble blown
by stellar winds.  As there is a massive molecular cloud to the west
of M\,17, we expect the hot gas to expand more rapidly to the east;
this flow of hot gas to the east has then blown the blister-like
bubble seen in optical images.

Four additional point sources are detected in the field around M\,17.
The brightest of these point sources lies to the south of the X-ray
peak and has been previously designated 1WGA~J1820.6$-$1615 in the WGA
Catalog of {\it ROSAT} Point Sources \citep{WGA94}.  This point source
is coincident with OI~352 \citep{OI76}, an O8 star on the southern
edge of the open cluster in M\,17.  The other three point sources lie
on the northern edge of the emission nebula.  These X-ray point
sources, designated as 1WGA~J1820.8$-$1603, 1WGA~J1821.0$-$1600, and
1WGA~J1820.8$-$1556, are coincident with stars GSC\footnote{The Guide
Star Catalog-I was produced at the Space Telescope Science Institute
under U.S. Government grant. These data are based on photographic data
obtained using the Oschin Schmidt Telescope on Palomar Mountain and
the UK Schmidt Telescope.}~06265$-$01977, SAO~161369 (a known O5
star), and GSC~06265$-$01808, respectively.  These point sources are
marked in Figure~\ref{fig:XrayOpt}a.

\subsection{X-Ray Spectra}
\label{sec:XraySpec}

\begin{figure}
\begin{center}
\epsscale{1.0}
\plotone{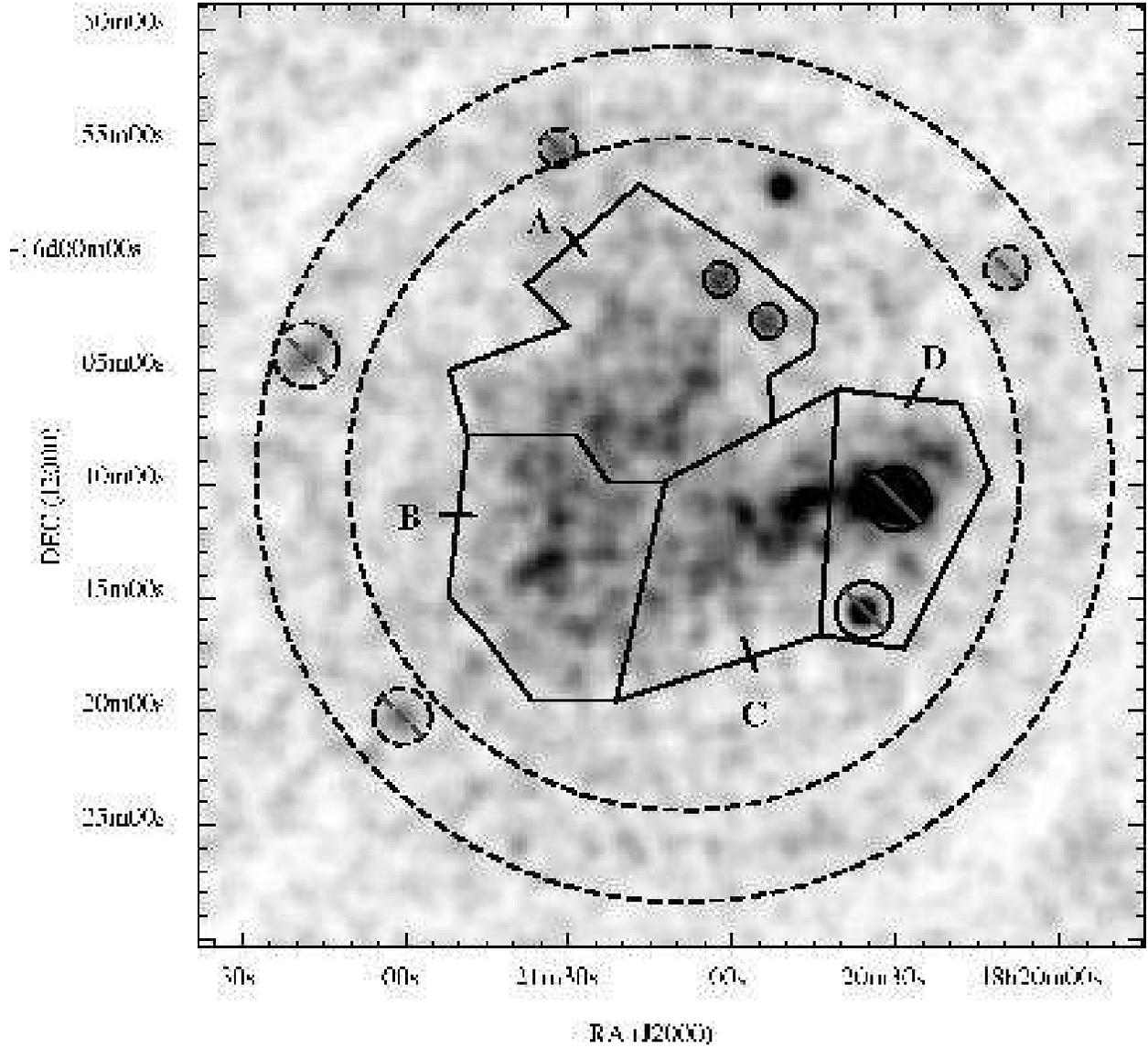}
\caption{{\it ROSAT} PSPC X-ray emission image overlaid with the
source regions used to extract and analyze the spectral properties of
the diffuse X-ray emission.  The source regions are A, B, C, and D.
The region used for background subtracted is indicated by the dashed
annulus.  Potential point-sources excluded from the source and
background regions are drawn as slashed circles.
\label{fig:XrayReg}}
\end{center}
\end{figure}

In order to examine the spectra of the diffuse X-ray emission, we
first excluded the point sources found in \S\ref{sec:XrayDist}.  Then,
noting that M\,17 is on the edge of a dense molecular cloud, we have
sectioned the nebula into four regions to account for anticipated
changes in the foreground absorption column density.  These regions
have been labeled A, B, C, and D and are displayed in
Figure~\ref{fig:XrayReg}.  Additionally, we have selected a background
annulus around the superbubble, as indicated in
Figure~\ref{fig:XrayReg}.  The background-subtracted spectra were then
extracted from the PSPC event files.

The observed X-ray spectra of the superbubble is a convolution of
several factors: the intrinsic X-ray spectrum of the superbubble, the
intervening interstellar absorption, and the PSPC response function.
Because the interstellar absorption and the PSPC response function are
dependent on photon energy, we must assume models of the intrinsic
X-ray spectrum and the interstellar absorption to make the problem
tractable.  As the X-ray emission from the superbubble interior
appears largely diffuse, we have used the \citet{RS77} thermal plasma
emission model to describe the intrinsic X-ray spectra of the
superbubble and the \citet{Morr83} effective absorption cross-section
per hydrogen atom for the foreground absorption, assuming solar
abundances for both the emitting and absorbing materials.  We then
simulated the observed spectrum, combining the assumed models for the
intrinsic spectrum and the interstellar absorption with the response
function of the PSPC.  The best-fit spectrum is found by varying
parameters and comparing $\chi^{2}$ for the simulated and observed
spectra.

\begin{figure}
\begin{center}
\epsscale{1.0}
\plotone{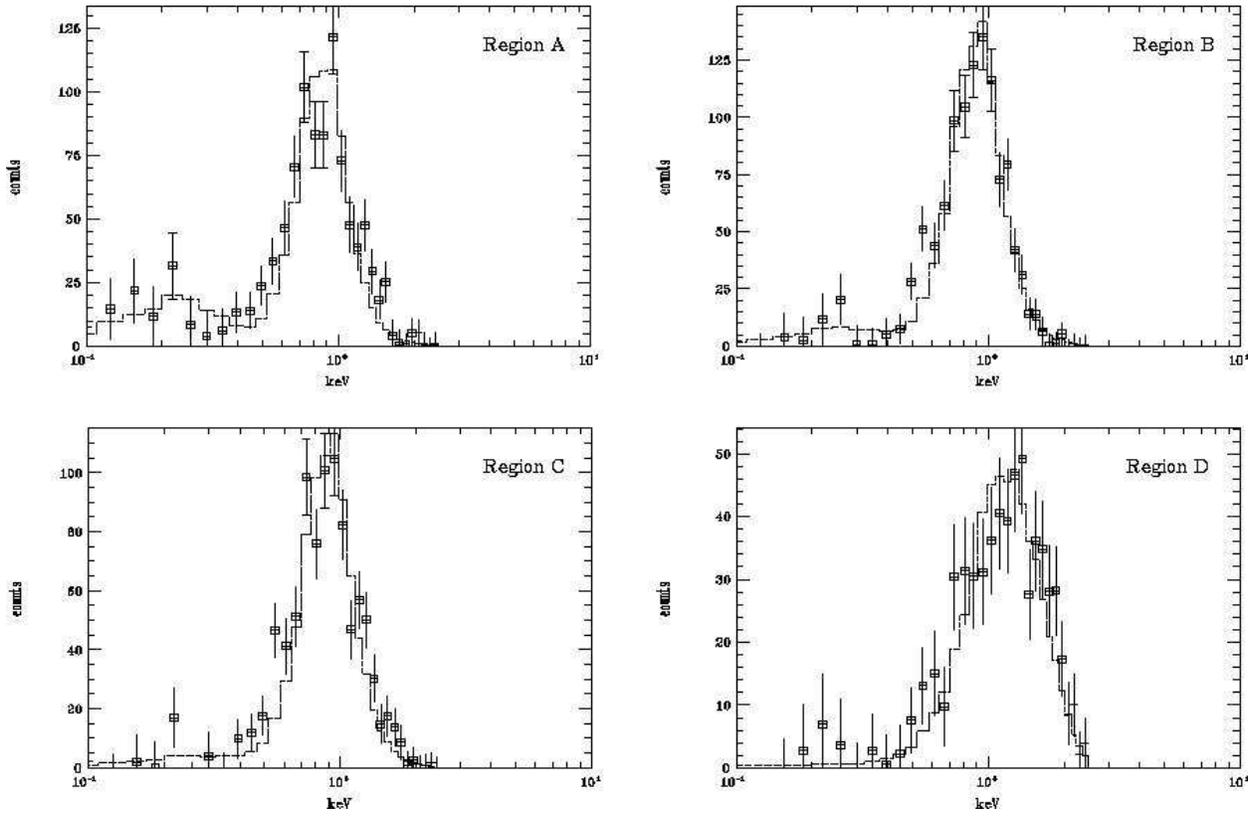}
\caption{These {\it ROSAT} PSPC count plots show the X-ray spectrum
extracted from each source region and the best thermal plasma model.
\label{fig:XraySpec}}
\end{center}
\end{figure}

\begin{figure}
\begin{center}
\epsscale{1.0}
\plotone{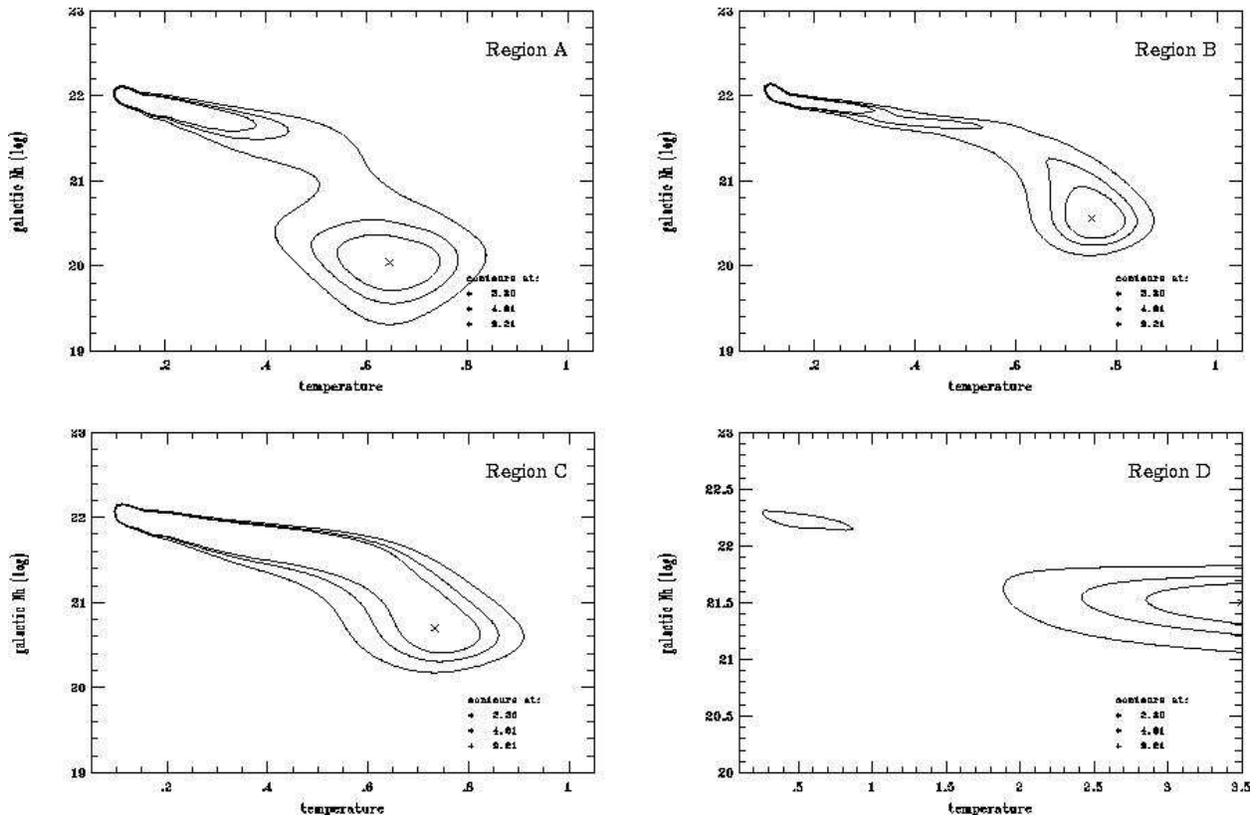}
\caption{$\chi^{2}$ plots for the X-ray spectral model fits to each
source region.  Confidence levels at 99\%, 90\%, and 68\% are
indicated as contours and the best fit location is marked with a
$\times$.
\label{fig:XrayGrid}}
\end{center}
\end{figure}

We performed a $\chi^{2}$ grid search of simulated spectral fits to
determine the best-fit levels for the thermal plasma temperature,
$kT$, and absorption column density, $N_{\rm H}$.  Plots of the best
fits to the X-ray spectra are shown in Figure~\ref{fig:XraySpec}, and
$\chi^{2}$ plots are presented in Figure~\ref{fig:XrayGrid}.  The
$\chi^{2}$ plots of regions A, B, and C indicate that the X-ray
emission can be fit by either higher temperature plasma,
$\sim$0.7~keV, with lower absorption column density,
$\sim$10$^{20-21}$~cm$^{-2}$, or lower temperature plasma,
$\sim$0.2~keV, with higher absorption column density,
$\sim$10$^{22}$~cm$^{-2}$.  This is a common problem for PSPC spectra
with a limited number of counts because of the poor spectral
resolution and soft energy coverage of the PSPC.  The best model fits
favor the higher temperature plasma and lower absorption column
density solution.  Indeed, the detection of soft X-ray emission below
0.5~keV indicates that the solution with $N_{\rm
H}$$\sim$10$^{22}$~cm$^{-2}$ cannot be valid, as such a solution
predicts no significant soft X-ray emission should be detected.
Furthermore, the high absorption column density solution predicts a
foreground absorption column density for M\,17 equal to the total
Galactic \ion{H}{1} column density along the line of sight \citep{DL90}.  As
M\,17 is located in the plane of the Galaxy at $l=15^{\circ}03'$,
$b=-00^{\circ}40'$ and has a distance of 1.6~kpc, we do not expect the
majority of the Galactic \ion{H}{1} toward this direction to be located in
front of the superbubble.

From the model fits, we calculated the unabsorbed X-ray flux, and
therefore the X-ray luminosity, $L_{\rm X}$, of the diffuse X-ray
emission from each source region.  The normalization factor, $A$, of
the thermal plasma model is equal to $\int n_{e}n_{p}dV/4\pi D^2$,
where $n_{e}$ and $n_{p}$ are the electron and proton number
densities, $V$ is the volume of the superbubble, and $D$ is the
distance to the source.  Assuming a He:H number ratio of 1:10 and that
the X-ray emitting gas is completely ionized, we find $n_{e}\simeq 1.2
n_{p}$, and that the volume emission measure can be expressed as
$<n_{e}^{2}>fV$, where $f$ is the volume filling factor.  We have used
the diameter of the superbubble, $\sim$10--12~pc, as the depth of the
X-ray emitting gas in each source region.  We determined the volume,
$V$, of each source region by multiplying the surface area of the
region by the depth and a geometric correction factor of 2/3
(approximated by the volume ratio of a sphere to a cylinder).  Taking
the volume filling factor to be $f$=0.5, we then calculated rms
$n_{e}$ in each region.  The best-fit values of $kT$, $A$, $N_{\rm
H}$, $L_{\rm X}$ and rms $n_{e}$ are given in
Table~\ref{tbl:XSpecFit}.  Note that the model fit to region
D did not converge because it contains a large number of
unresolved stellar sources as well as diffuse emission; only
approximate values for the X-ray luminosity and absorption column
density are given.  See \citet{Townsley03} for a detailed analysis of
the {\it Chandra} observations of region D.

Combining our results from each of the source regions, we find a total
diffuse X-ray luminosity of $\sim$2.5$\times$10$^{33}$~ergs~s$^{-1}$
in the {\it ROSAT} PSPC 0.1--2.4~keV band, a mean characteristic
temperature of $kT\sim$~0.72~keV or $T\sim$8.5$\times$10$^6$~K, a mean
electron density of $\sim$0.09~cm$^{-3}$, and a total hot gas mass
of $\sim$1~M$_{\odot}$.  We have calculated the total thermal energy
in hot, shocked wind component of the superbubble to be $E_{\rm
th}\sim$~1$\times$10$^{48}$~ergs with a cooling timescale of
$t_{T}\sim$~40~Myr.  Given the stellar age of the open cluster,
$\sim$1~Myr, we do not expect significant radiative cooling to have
occurred.  As a rough check, we multiply the current X-ray luminosity
by the age of the cluster and find that the total energy radiated away
by the X-ray emission is $\lesssim$10\% of the total thermal energy.

\section{Discussion}
\label{sec:Discuss}

\subsection{Comparisons with Model Expectations}
\label{sec:CompMod}

We now compare the observed physical properties of M\,17 determined
above to theoretical calculations from basic wind-blown bubble models.
Although the M\,17 superbubble is in an inhomogeneous ambient medium
with a significant density gradient and the cluster is off-centered (a
more complex scenario than is considered in basic bubble models), if
the superbubble structure is indeed governed by the physical processes
prescribed by these models, we expect the properties of the diffuse
X-ray emission to agree with predictions within similar orders of
magnitude.  We first consider the wind-blown bubble model of
\citet{Weaver77} and will later consider a wind-blown bubble model
without heat conduction.

\subsubsection{A Bubble with Heat Conduction}

In the Weaver et al.\ model, heat conduction and mass evaporation act
across the boundary between the hot interior gas and the nebular shell
to lower the temperature and raise the density of the bubble interior.
The temperature and electron density profiles of such a bubble have
been calculated by \citet{Weaver77}, and the X-ray luminosities of
such bubbles can be determined using two methods outlined by
\citet{Chu95}.  In the first method, we derive the expected X-ray
luminosity from the observed physical properties of the gas in the
10$^4$~K ionized shell of swept-up ISM.  In the second method, we use
the spectral types of massive stars in M\,17 to estimate the combined
mechanical luminosity of the stellar winds and then derive the
expected X-ray luminosity.  These two methods use independent input
parameters and thus allow us to check the consistency of the
pressure-driven bubble model in addition to comparing the expected and
observed X-ray luminosities.

\subsubsubsection{X-Ray Luminosity Method 1: The Ionized Shell}

The expected X-ray luminosity in the {\it ROSAT} band of 0.1--2.4~keV
for the \citet{Weaver77} wind-blown bubble model has been given by
\citet{Chu95},
\begin{equation}
L_{\rm X} = (8.2 \times 10^{27}~{\rm ergs~s}^{-1}) \xi I(\tau)n_{0}^{10/7}R_{\rm pc}^{17/7}V_{\rm km/s}^{16/7},
\end{equation}
where $\xi$ is the metallicity relative to the solar value and in this
case we assume a value of unity, $I(\tau)$ is a dimensionless integral
of value $\sim$2, $n_{0}$ is the number density of the ambient medium
in cm$^{-3}$, $R_{\rm pc}$ is the radius of the superbubble in pc,
$V_{\rm km/s}$ is the expansion velocity of the superbubble in units
of \kms.  The ambient density $n_{0}$ cannot be measured directly, but
assuming that the ram pressure of the expanding shell is equal to the
thermal pressure of the ionized superbubble shell, the relation
between the ambient density and the density of the ionized shell is
given by
\begin{equation}
n_{0} = (9/7)n_{\rm i}kT_{\rm i}/(\mu_{a}V_{\rm exp}^{2}),
\end{equation}
where $n_{\rm i}$ is the electron number density in the ionized shell,
$T_{\rm i}\sim$~10$^4$~K is the electron temperature in the ionized
shell, $V_{\rm exp}$ is the expansion velocity of the bubble, and
$\mu_{a}=(14/11)m_{\rm H}$ \citep{Weaver77,ChuMacLow90}.  Adopting a
mean electron density of $n_{\rm i}\sim$~300~cm$^{-3}$
\citep{Felli84} and an observed $V_{\rm exp}\sim$~25~\kms\
\citep{Clayton85}, we calculated an ambient density of
$n_{0}\sim$~40~cm$^{-3}$.  Given the superbubble radius of
5--6~pc, we have determined an expected X-ray luminosity of
$\sim$3$\times$10$^{35}$~ergs~s$^{-1}$.

\subsubsubsection{X-Ray Luminosity Method 2: Wind Luminosity from OB Stars}
\label{sec:WindLum}

We can also calculate the expected X-ray luminosity in an
energy-conserving, wind-blown bubble by the following equation from
\citet{Chu95},
\begin{equation}
L_{\rm X} = (1.1 \times 10^{35}~{\rm ergs~s}^{-1}) \xi I(\tau)L_{37}^{33/35}n_{0}^{17/35}t_{\rm Myr}^{19/35},
\end{equation}
where $L_{37}$ is the mechanical luminosity of the stellar winds in
units of 10$^{37}$~ergs~s$^{-1}$, and $t_{\rm Myr}$ is the age of the
bubble in Myr.  To remain independent of Method 1, we do not use the
value of $n_{0}$ determined for that method.  Rather, we use the
following relations between ambient density, radius, wind luminosity,
bubble age, and expansion velocity,
\begin{equation}
n_{0} = (1.3 \times 10^{8}~{\rm cm}^{-3}) L_{37}t_{\rm Myr}^{3}R_{\rm pc}^{-5},
\end{equation}
\begin{equation}
t_{\rm Myr} = (0.59 {\rm Myr})R_{\rm pc}/V_{\rm km/s},
\end{equation}
\citep{Weaver77,Chu95}.  We again take the expansion velocity to be
$\sim$25~\kms\ \citep{Clayton85} and the radius to be 5--6~pc
and derive a bubble age of $\sim$0.13~Myr.  

To determine the wind luminosity of M\,17, we examined its massive
stellar content.  \citet{Hanson97} identified nine O stars and four
late-O/early-B stars in the open cluster.  Using the spectral types of
these massive stars, we have estimated their terminal stellar wind
velocities, effective temperatures, and luminosities based on the
stellar parameters given by \citet{Prinja90} and \citet{VGS96}.  We
then calculated the mass-loss rates for the OB stars in M\,17 by
utilizing the empirically derived relationship between effective
temperature, luminosity, and mass-loss rate of \citet{dJ88}.
Table~\ref{tbl:StarPar} lists the detected OB stars, their spectral
types as determined from optical and K-band observations, their
terminal wind velocities $V_{\infty}$, stellar effective temperatures
$T_{\rm eff}$, stellar luminosities $L$, and their mass-loss rates
$\dot{M}$.  We calculated the total mechanical luminosity of the
stellar winds,
\begin{equation}
L_{\rm w} = \Sigma (1/2) \dot{M} V_{\infty}^{2},
\end{equation}
from the OB stars to be $\sim$1$\times$10$^{37}$~ergs~s$^{-1}$.  As
noted by \citet{Felli84}, the identified OB stars can approximately
account for the ionization of the emission nebula; we therefore expect
our calculated wind mechanical luminosity to be reasonably complete as
well.  Assuming a relatively constant mechanical luminosity, we find a
total energy deposited by stellar winds of
$\sim$4$\times$10$^{49}$~ergs over the life of the bubble.  In
addition, this value of the wind mechanical luminosity gives an
ambient density of $\sim$60~cm$^{-3}$ and an expected X-ray luminosity
of $\sim$5$\times$10$^{35}$~ergs~s$^{-1}$.  This X-ray luminosity
value is consistent to within a factor of two with the value found by
Method 1.

Although the two methods of determining the expected X-ray luminosity
are consistent with each other, they do not agree with the X-ray
luminosity derived from the PSPC observation.  The observed X-ray
luminosity is $\sim$100--200 times lower than expected from Weaver et
al.'s bubble model.  It is possible that stellar winds are clumpy, as
suggested by \citet{Moffat94}, then the conventionally derived mass
loss rates would be reduced by a factor of $\ge$3.  Even using the
reduced mass loss rates, the expected X-ray luminosity is more than 40
times too high.  The observed temperature, density, and surface
brightness of the hot gas in M\,17 do not agree with the model
expectations, either.  The physical conditions of the shocked stellar
winds in Weaver et al.'s model are heavily modified by heat conduction
and the hot gas mass is dominated by the nebular mass evaporated
across the interface.  The predicted temperature is
5.0--5.6$\times$10$^6$ K near the center and decreases outward, the
predicted density is 0.2--0.4 cm$^{-3}$ near the center and increases
outward, and the X-ray surface brightness is expected to show
limb-brightening.  Compared with observed properties, the expected
temperature is too low, density is too high, and the X-ray morphology
is inconsistent.  The disagreements between observations and model
expectations suggest that heat conduction may not play a dominant role
in determining the physical conditions inside this superbubble.

Heat conduction can be suppressed by the presence of magnetic fields
\citep{Soker94,BL88}.  The magnetic field strength in M\,17 has been
measured via the \ion{H}{1} Zeeman effect to be 100--550~$\mu$G,
peaking near the interface between the \ion{H}{2} region and the
molecular cloud M17SW \citep{Brogan99}.  Assuming a comparable
magnetic field strength in the swept-up 10$^4$~K shell, we find the
Alfv\'en speed to be 10--60~\kms\ which is comparable to or much
greater than the isothermal sound velocity of the 10$^4$~K gas,
10~\kms for H atoms.  In addition, the magnetic field strength and
isothermal sound velocity indicate a gyro-radius of $\lesssim$10~km
for protons in the swept-up shell.  This suggests that the protons in
the 10$^4$~K gas will be unable to escape the magnetic field and
diffuse into the interior of the superbubble, inhibiting heat
conduction and mass evaporation between the hot interior and the cool
shell of the bubble.

\subsubsection{A Bubble without Heat Conduction}

We now turn our consideration to a wind-blown bubble without heat
conduction.  The X-ray emission of a bubble interior depends on both
the temperature and the amount of hot gas.  We will first compare the
plasma temperature expected from the shocked stellar winds to the
observed hot gas temperature.  Using the combined stellar winds
mass-loss rate of 4.3$\times$10$^{-6}$~M$_\odot$~yr$^{-1}$ (summed
from Table~\ref{tbl:StarPar}) and the integrated wind mechanical
luminosity $L_{\rm w}$ of 1$\times$10$^{37}$~ergs~s$^{-1}$ as
calculated in \S\ref{sec:WindLum}, we derive an rms terminal wind
velocity of $V_\infty \sim$~2700~\kms.  The post-shock temperature of
the combined stellar winds is therefore expected to be
$\sim$8$\times$10$^7$~K.  This temperature is an order of magnitude
higher than that indicated by PSPC observations; to lower it to the
observed temperature of $\sim$8.6$\times$10$^6$~K requires the mixing
in of cold nebular mass that is nearly 10 times the mass of the
combined stellar winds.  This mixing may be provided by turbulent
instabilities at the interface between the shocked fast winds and the
cold nebular shell \citep[e.g.,][]{SS98} or through the hydronamic
ablation of clumps of cold nebular material distributed within the
hot bubble interior \citep{Pitt01}. 

Assuming that mixing has taken place, we next determine the hot gas
mass expected as a result of mixing and compare it to the observed
value (\S\ref{sec:XraySpec}).  Given the dynamical age of the
superbubble, 0.13~Myr, a total stellar wind mass of
$\sim$0.56~M$_\odot$ has been injected to the superbubble interior,
and the expected total mass of the hot gas will be 5--6~M$_\odot$.
This is significantly greater than the observed value of
$\sim$1~M$_\odot$.  This discrepancy can be reduced if we again
consider the possibility of clumpy stellar winds \citep{Moffat94}.
With the mass-loss rates reduced by a factor of $\ge$3, the expected
hot gas mass is $\sim$2~M$_{\odot}$, which would be in remarkable
agreement with the observed value.

We summarize the comparison between the observed X-ray emission and
the various models in Table~\ref{tbl:XrayProp}, which lists the
observed X-ray luminosity and hot gas temperature and mass as well as
those expected from models with and without heat conduction for both
homogeneous winds and clumpy winds.  It is clear that bubble models
with heat conduction have the largest discrepancies from the
observations.  The best agreement with observed properties is from
bubble models without heat conduction but allowing dynamical mixing of
cold nebular material with the hot gas.  For models either with or
without heat conduction, clumpy winds with reduced mass loss rates are
needed to minimize the discrepancy between model expectations and
observations.

\subsection{Comparisons with Other Wind-Blown Bubbles}
\label{sec:CompBubb}

Diffuse X-ray emission has been previously detected from other types
of wind-blown bubbles, including planetary nebulae (PNe) and
circumstellar bubbles blown by Wolf-Rayet (WR) stars.  The X-ray
emission from these circumstellar bubbles is qualitatively and
quantitatively different from that of M\,17.  \citet{CGG03} find that
the X-ray emission from PNe and WR bubbles shows a limb-brightened
spatial distribution, in sharp contrast to the centrally-filled
spatial distribution in M\,17 as described in \S\ref{sec:XrayDist}.
Further, \citet{CGG03} note that PNe and WR bubbles exhibit hot gas
temperatures of 1--3$\times$10$^6$~K and electron densities of
10--100~cm$^{-3}$, while the hot interior gas of M\,17 exhibits a
temperature of 8.5$\times$10$^6$~K and a substantially lower electron
density of $\sim$0.09~cm$^{-3}$.  

The comparisons of morphology and temperature between the M\,17
superbubble and small circumstellar bubbles show fundamental
differences.  The limb-brightened X-ray spatial distribution, low
temperatures, and high electron densities of PNe and WR bubbles are
qualitatively consistent with a hypothesis of heat conduction and mass
evaporation occurring between the hot gas interior and the swept-up
shell.  However, the observed X-ray luminosities for PNe and WR
bubbles are both significantly lower (10--100 times) than predicted by
bubble models with heat conduction \citep{Chu01,Wrigge94,Wrigge99}.
It is possible that in these wind-blown bubbles, heat conduction has
also been suppressed and that dynamical mixing, which allows a lower
mass injection rate, occurs at the interface between the hot gas
interior and the cool nebular shell.  Exploring this question will
require magnetic field measurements of PNe and WR bubbles.

\section{Summary}
\label{sec:Summary}

We have presented analysis of a {\it ROSAT} observation of the
emission nebula M\,17.  The blister-like morphology seen in the
optical images indicates that it is a superbubble blown by the winds
of its OB stars in an inhomogeneous ISM.  With a stellar age of
$\sim$1~Myr, M\,17 must be a young quiescent superbubble without any
supernova heating.  Diffuse X-ray emission is detected from M\,17 and
is confined within the optical shell.  This suggests the presence of
hot 10$^6$--10$^8$~K gas in the interior of M\,17, as is expected in a
wind-blown bubble.  Analysis of the diffuse X-ray emission indicates a
characteristic gas temperature $\sim$8.5~$\times$~10$^6$~K with a mean
electron number density of 0.09~cm$^{-3}$.

We have considered bubble models with and without heat conduction and
found that those with heat conduction overpredict the X-ray luminosity
by two orders of magnitude.  Furthermore, the magnetic field measured
in M\,17 is large enough to suppress heat conduction and associated
mass evaporation.  Bubble models without heat conduction overestimate
the hot gas temperature unless mixing with cold nebular gas has
occurred.  If nebular gas can be dynamically mixed into the hot bubble
interior and if the stellar winds are clumpy with a lower mass-loss
rate, the X-ray morphology and luminosity, and hot gas temperature and
mass can be reasonably reproduced.

M\,17 provides us a rare opportunity to probe the physical conditions
of hot gas energized solely by shocked stellar winds in a quiescent
superbubble.  While we have learned much from the current analysis,
our model considerations were performed on a very basic level.  More
robust models are needed to accurately describe the evolution of a
superbubble in a medium with a large-scale density gradient,
small-scale clumpiness, and a strong magnetic field.

\acknowledgments{We would like to thank the anonymous referee for the 
stimulating comments with have helped us to improve this paper.  This
research has made use of data obtained through the High Energy
Astrophysics Science Archive Research Center Online Service, provided
by the NASA/Goddard Space Flight Center.}

\clearpage

\clearpage

\clearpage

\clearpage

\clearpage

\begin{table}
\begin{center}
\caption{M\,17 X-Ray Spectral Fits \label{tbl:XSpecFit}}
\begin{tabular}{lccccc}
\tableline\tableline
Region & $kT$  & $A$ & log $N_{\rm H}$ & $L_{\rm X}$\tablenotemark{a} & rms $n_{e}$ \\
       & [keV] & [cm$^{-5}$] & [cm$^{-2}$] & [ergs~s$^{-1}$] & [cm$^{-3}$] \\
\tableline
A  & 0.65 & 4.3$\times$10$^{10}$ & 20.04 & 4.4$\times$10$^{32}$ & 6.5$\times$10$^{-2}$ \\
B & 0.75 & 7.0$\times$10$^{10}$ & 20.57 & 6.3$\times$10$^{32}$ & 9.1$\times$10$^{-2}$ \\
C  & 0.73 & 5.3$\times$10$^{10}$ & 20.70 & 5.2$\times$10$^{32}$ & 9.3$\times$10$^{-2}$ \\
D\tablenotemark{b}   & --  & -- & 21.5  & 1$\times$10$^{33}$ & -- \\
\tableline
\end{tabular}
\tablenotetext{a}{In the 0.1--2.4~keV energy band}
\tablenotetext{b}{Fit did not converge, values approximated}
\end{center}
\end{table}

\clearpage
\begin{table}
\begin{center}
\caption{OB Stars in M\,17 and their Stellar Wind Parameters \label{tbl:StarPar}}
\begin{tabular}{lcccccc}
\tableline\tableline
Star    & Optical & K-Band  & $V_{\infty}$\tablenotemark{d} & 
$T_{\rm eff}$\tablenotemark{e} & log $L/L_{\odot}$\tablenotemark{e} & 
log $\dot{M}$\tablenotemark{f} \\
Number\tablenotemark{a} & Spectral Type\tablenotemark{b} & 
Spectral Type\tablenotemark{b} & [km~s$^{-1}$] & 
[K] &  & [$M_{\odot}$~yr$^{-1}$] \\
\tableline
B98     & O9V     & kO9--B1 & 1500 & 36000 & 5.1 & $-$6.8 \\
B111    & O5V     & kO3--O4 & 2900 & 46000 & 5.7 & $-$6.2 \\
B137    & ...     & kO3--O4 & 3100 & 50000 & 6.0 & $-$6.0 \\
B164    & O7--O8V & kO7--O8 & 2000 & 40000 & 5.3 & $-$6.6 \\
B174    & ...     & kO3--O6 & 2900 & 47000 & 5.8 & $-$6.2 \\
B181    & ...     & kO9--B2 & 1500 & 33000 & 4.7 & $-$7.2 \\
B189    & O5V     & kO3--O4 & 2900 & 46000 & 5.7 & $-$6.2 \\
B243    & early B & \tablenotemark{c} & 500 & 22000 & 3.3 & $-$11.1 \\
B260    & O7--O8V & kO3--O4 & 2000 & 40000 & 5.3 & $-$6.6 \\
B268    & B2      & \tablenotemark{c} & 500 & 22000 & 3.3 & $-$11.1 \\
B289    & O9.5V   & \tablenotemark{c} & 1500 & 35000 & 5.0 & $-$6.9 \\
B311    & ...     & kO9--B2 & 1500 & 33000 & 4.7 & $-$7.2 \\
OI~345  & O6V     & kO5--O6 & 2600 & 44000 & 5.6 & $-$6.4 \\
\tableline
\end{tabular}
\tablenotetext{a}{From \citet{Bumgardner92}, except OI~345 from \citet{OI76}}
\tablenotetext{b}{From \citet{Hanson97}, except OI~345 from \citet{Crampton78}}
\tablenotetext{c}{Shows excess K-band emission, most likely a young stellar object.}
\tablenotetext{d}{Adopted from \citet{Prinja90}.}
\tablenotetext{e}{Adopted from \citet{VGS96}.}
\tablenotetext{f}{From empirical relationship of \citet{dJ88}.}
\end{center}
\end{table}

\clearpage
\begin{table}
\begin{center}
\caption{M\,17 Observed and Model X-ray Properties \label{tbl:XrayProp}}
\begin{tabular}{lccc}
\tableline\tableline
         & $L_{\rm X}$                & Temperature & Hot Gas Mass \\
         & [10$^{33}$~ergs~s$^{-1}$]  & [10$^6$~K]  & [M$_{\odot}$] \\
\tableline
Observed & 2.5                        & 8.5         & 1 \\
\\
Model: Heat Conduction \\
~~~Method 1\tablenotemark{a} & 3.1$\times$10$^2$ & 5.0 & $\gtrsim$4 \\
~~~Method 2 \\
~~~~~~Homogeneous Winds\tablenotemark{a} & 5.2$\times$10$^2$ & 5.6 & $\gtrsim$9 \\
~~~~~~Clumpy Winds\tablenotemark{a} & 1.1$\times$10$^2$ & 4.1 & $\gtrsim$4 \\
\\
Model: No Heat Conduction \\
~~~Without Dynamical Mixing \\
~~~~~~Homogeneous Winds & 0.2        & 80         & 0.5 \\
~~~~~~Clumpy Winds      & 0.03       & 80         & 0.2 \\
~~~With Dynamical Mixing \\
~~~~~~Homogeneous Winds & 75         & 8.5        & 5--6 \\ 
~~~~~~Clumpy Winds      & 10         & 8.5        & 2 \\
\tableline
\end{tabular}
\tablenotetext{a}{Central temperature as derived from the \citet{Weaver77} model.}
\end{center}
\end{table}

\end{document}